18/12/13

# Measurement, correction and implications of the intrinsic error fields on MAST


A. Kirk, Yueqiang Liu, R. Martin, G. Cunningham, D. Howell and the MAST team

EURATOM/CCFE Fusion Association, Culham Science Centre, Abingdon, Oxon, OX14 3DB, UK



**Abstract**

The misalignment of field coils in tokamaks can lead to toroidal asymmetries in the magnetic field, which are known as intrinsic error fields. These error fields often lead to the formation of locked modes in the plasma, which limit the lowest density that is achievable. The intrinsic error fields on MAST have been determined by the direct measurement of the toroidal asymmetry of the fields from these coils and have been parameterised in terms of distortions to the coils. The error fields are corrected using error field correction coils, where the optimum correction is found by determining the current required to ensure that the discharge is furthest from the onset of a locked mode. These empirically derived corrections have been compared with the known coil distortions. In the vacuum approximation there is a factor of ~ 3 difference between the predicted and empirically determined correction. When the plasma response is included better agreement is obtained, but there are still some cases where the agreement is not good, which suggests that other effects such as the non-linear coupling of the error field to the plasma are important.




## 1.    Introduction

In a tokamak misalignment of field coils, coil connections and even the finite number of discretised coils used to produce the toroidal field lead to asymmetries in the magnetic field known as intrinsic error fields.  While it is possible to minimise the size of these error fields, by careful design, manufacture and installation, it is not possible to remove the source of them completely.  Error fields have been observed to have a detrimental effect on plasmas on many machines [1]-[7]. For example, it is known that error field as small as ~ $10^{-4}$ of the toroidal magnetic field can induce non-rotating, or locked, tearing modes in the plasma [8][9].  To reduce the effect of the intrinsic error fields an additional set of error field correction coils are used.  These coils are designed to create non-axi-symmetric fields inside the tokamak which oppose the intrinsic error field.  Application of these error field correction coils (EFCCs) allow the machines to operate in previously inaccessible regimes, typically at low density.  However, these coils rarely correct the intrinsic error field completely, they typically just minimise it at a certain plasma location, typically the q=2 surface.  For example, EFCCs installed on MAST, which produce a magnetic perturbation with a toroidal mode number (n) of 1 have been used to minimise the resonant component of the error field on the q=2 surface ($B_{21}$) [7].  The application of such a correction has allowed discharges with a 30 % lower plasma density to be produced without termination due to a locked mode.

In recent years, non-axisymmetric fields with n = 1-6 are routinely applied to tokamaks in order to modify ELM behaviour in H-mode plasmas. This technique of Resonant Magnetic Perturbations (RMPs) has been employed on DIII-D **[10][11]** and



KSTAR **[12]**, where complete ELM suppression has been possible, and on JET **[13]**, MAST **[14]** and ASDEX Upgrade **[15]** where ELM mitigation has been obtained. In these experiments it is often observed that the ELM suppression and/or mitigation depends on the phase of the applied perturbation with respect to the machine coordinate system [16]. This is explained as being due to the interaction between the applied RMP field and the intrinsic error fields and the applied error field corrections. Hence in order to fully understand the RMP experiments it is important to have a good understanding of the intrinsic error fields in a device.

In this paper, measurements of the dominant intrinsic error fields in MAST are presented in section 2 while in section 3 the error field correction coils are described. In section 4 the experimentally determined error field correction will be discussed and Section 5 presents a comparison of the measured intrinsic error field and the determined correction using both vacuum and plasma response modelling before concluding in Section 6 with an outlook for further studies.

## 2. *Measurement of the intrinsic error fields due to the poloidal field coils*

A number of other tokamaks have previously described direct magnetic measurement of the intrinsic error fields due to coil misalignments (see for example [17] and references therein). Previous error field experiments on MAST [7] suggested that the main poloidal field coils, called P4 and P5 (see Figure 1a for their location), were the dominant contributors to the intrinsic error field. Each poloidal field coil is an up-down symmetric pair (e.g. P4U and P4L etc). Further evidence comes from the fact that it was known that the manufacturing process used for these coils led to a certain amount of distortion. In order



to measure the error field due to these coils  a scheme was devised to accurately measure the position of the P4 and P5 coils, and the radial and vertical fields they produce.

## 2.1 Measurement of the PF coil positions and fields

A clamp ring was devised that could be attached either to the top or bottom of the MAST centre tube. An accurately machined arm was then fixed to the clamp ring and gauges attached to the arm were then used to measure the position of the coils.  The position of the clamp ring was measured with respect to the end plates of the MAST vessel. This method only allowed a determination of the position of the coil case, not the copper turns of the coil inside the case.  To make the magnetic field measurements, three high accuracy Hall probes were installed on a mounting plate attached to the measuring arm. One Hall probe was used as a reference while the other two where used to make measurements at differing toroidal locations of the radial and vertical field.  Vertical field measurements where made at twenty-four toroidal angles, every 15°. The vertical and radial location of the measurement positions is shown by the dots in Figure 1a, whilst the toroidal location is shown in Figure 1b.  The radial field measurements where made at twelve toroidal angels, every 30°. For the vertical field measurements the reference probe was aligned vertically, for radial measurements it was aligned radially. The (r, z) location for each detector was accurately measured relative to the vessel end plates and hence the coil cases. .

A single low ripple constant current source power supply was connected to each coil in turn and a stationary flat top current of ~3.5 kA for 2 s was driven through the 23 turns of each PF coil (i.e. P4U, P4L, P5U, P5L). The radial and vertical components of the



field from that coil was then determined as a function of toroidal location. The long flattop and low ripple of the power supply meant that the effect of induced currents could be ignored.

The sample rate of the Hall probe system was 10 Hz meaning that 20 measurements were made per position during the 2s current flattop and a number of repeats (shots) were performed, which allowed any outliers to be removed. The data from different shots were scaled, using the reference Hall probe signal and a cross calibration between the radial and vertical field measurements performed using the measured coil current. The result, shown in Figure 2 and Figure 3 for the P4 and P5 coils respectively, was a measurement of the mean and standard deviation of the radial and field components as a function of toroidal angle. As can be seen from the toroidal profiles, the radial and vertical fields have a complex structure, which require more than a single harmonic component to describe them (i.e. they do not have a simple $\cos\phi$ dependence).

## 2.2 Modelling of the PF coils

Each turn of the PF coil has been modelled as a single closed loop, which is parameterised in terms of the toroidal angle ($\phi$). This loop is then modified using various vector transformations such that the magnetic field that it would produce is a good fit to the measured field. A shift or tilt of the coil produces a toroidal n=1 asymmetry, while a distortion from a flat circle, either in terms of an elliptic or out of plane bowing, produces an n=2 asymmetry. The vector representation **r**, for an elliptical coil in the XY-plane can simply be written as



$$r_x = (r_o + \varepsilon)\sin(\phi) \qquad\qquad (1)$$

$$r_y = (r_o - \varepsilon)\cos(\phi) \qquad\qquad (2)$$

$$r_z = 0 \qquad\qquad (3)$$

where $r_o$ is the mean radius of the filament and $\varepsilon$ the deformation from a circle. This ellipse is only deformed in the direction of the x/y-axes, using a vector rotation about the z-axis, an ellipse with any axis of deformation can be created.

$$\underline{\mathbf{r}} = \mathrm{Rot}_Z(\underline{\mathbf{r}}, \phi_\varepsilon) \qquad (4)$$

where $\phi_\varepsilon$ is the rotation angle. The filament can now be deformed into a saddle or bowed shape by setting $r_z$

$$r_z = \beta(\cos^2(\phi - \phi_\beta) - 0.5) + \delta z \qquad (5)$$

where $\phi_\beta$ is the angular offset from the x-axis for the coil bowing, $\beta$ the amount of deflection of the coil and $\delta z$ is the initial filament offset in the z-direction relative to the centre of the coil pack. A possible n=3 component has been included by adding a "Wobble term" of the form

$$r_z = r_z + \omega\cos(3\phi - \phi_\omega) \qquad (6)$$

Finally a tilt and shift of the coil is then allowed using

$$\underline{\mathbf{r}} = \mathrm{Rot}_Z(\mathrm{Rot}_y(\mathrm{Rot}_Z(\underline{\mathbf{r}}, -\phi_t), \lambda), \phi_t) \qquad (7)$$

$$\underline{\mathbf{r}} = \underline{\mathbf{r}} + \underline{\mathbf{r}}_s \qquad\qquad (8))$$

where $\phi_t$ is the angle between the axis of tilt and the y-axis, $\lambda$ is the angle of tilt and $\underline{\mathbf{r}}_s$ is the displacement of the coil from the origin.



The field due to a closed circular current carrying filament, is calculated using the Biot-Savart integral

$$B = I \oint \vec{r} \times \delta\ell / r^3 \qquad\qquad (9)$$

where I is the current in the coil. The filament is split into $n$ small line segments, with position vector $\mathbf{r}_i$ . The field due to the PF coils is then described by 12 parameters, giving the field at a particular detector location as

$$\mathrm{B_{measured}} = \mathbf{B}(I, \mathrm{r_x}, \mathrm{z_x}, \phi_x; \mathrm{r_0}, \mathrm{z_0}, \varepsilon, \beta, \lambda, \phi_\varepsilon, \phi_\beta, \phi_w \phi_t, x, y, \delta z).\mathbf{n} \qquad (11)$$

Where $\mathbf{n}$ is the vector orientation of the detector, I the coil current, $\mathrm{r_x}$, $\mathrm{z_x}$ , $\phi_x$ the coordinate of the detector, $\mathrm{r_0}$, $\mathrm{z_0}$ the nominal position of the PF coil and x,y the displacement of the coil. The ellipticity, bowing and tilt distortion ($\varepsilon$, $\beta$, $\lambda$) parameters and the corresponding angular positions of the axis of each distortion ($\phi_\varepsilon$ , $\phi_\beta$ , $\phi_w$ $\phi_t$) are described above. The parameters, shown in Table 1, are determined for each coil by performing a least squared fit to the radial and vertical field measured as a function of toroidal angle. The results of the fits are shown in Figure 2 and Figure 3. The parameterisation appears to be a good description of the measured values.

### 3. Error field correction coils on MAST

On MAST, the correction of the n=1 component of the intrinsic error field is normally attempted using four ex-vessel coils [7], which are arranged symmetrically around the outside of the vacuum vessel, with each coil spanning 83∘ toroidally (see Figure 1). Each coil consists of three turns and can carry a maximum current of 15 kA turns (kAt). In routine use the coils are arranged in two pairs, with opposite coils wired in series to



produce a non-axisymmetric magnetic field with an odd-n spectrum. The two pairs of correction coils are powered by independent power supplies, which allow an n = 1 field to be applied at an arbitrary toroidal phase. The coils are grouped in pairs which are wired in series and are referred to as 'EFCC_2_8' and 'EFCC_5_11' in this paper. EFCC_5_11 are the coils centred on sectors 5 and 11 (at 315 and 135 degrees respectively, see Figure 1b). EFCC_2_8 are the coils centred on sectors 2 and 8 (at 45 and 225 degrees respectively, see Figure 1b). The convention for the sign of current in the coils is that a positive current in EFCC_2 produces a radial field pointing inwards ($B_r<0$) at sector 2 and a field pointing outwards ($B_r>0$) at sector 8. Similarly a positive current in EFCC_5 produces a radial field pointing outwards ($B_r>0$) at sector 5 and a field pointing inwards ($B_r<0$) at sector 11.

It is also possible to attempt correction of the intrinsic error field using the internal ELM control coils [18] in an n=1 configuration. The ELM control coils are located inside the vacuum vessel, between the P4 and P5 coils (see Figure 1a) in two rows, with 12 coils in the lower row and 6 coils, located at the odd sectors, in the upper row (see Figure 1b). The configuration of the coils used to produce the n=1 correction is described in section 4.2.

## 4. Determination of optimised n=1 error field correction

The optimum error field correction on MAST is defined such as to maximise the distance from the onset of a locked mode. Two methods have been used on MAST to determine the onset criterion for locked modes and hence determine empirically the optimum error field correction: Either ramping the applied field at constant density or using a decreasing density at fixed applied field. The experiments were carried out in Ohmic, L mode, double



and single null plasmas. Plasmas with toroidal fields in the range (0.35–0.6) T, plasma currents in the range (450–750) kA and line averaged densities in the range (0.5–3.0)$\times 10^{19}$m$^{-3}$ were used. The normalised plasma pressure ($\beta_N$) was in the range 0.8 to 1.2 i.e. well away from any stability limits. Since it is believed that the source of the intrinsic field is due to the P4 and P5 coils the experiments have been performed at a range of current in P5 ($I_{P5}$) and P4 ($I_{P4}$) and in particular a range of $I_{P5}/I_{P4}$. Most plasma scenarios in MAST use a ratio $I_{P5}/I_{P4}$ in the range 0.4 – 0.7 but to improve the sensitivity of the measurements shots have been developed at low $I_{P5}/I_{P4} = 0.25$ and high $I_{P5}/I_{P4} = 1.8$ ratios.

### 4.1 Determination of the intrinsic n=1 error field correction using the EFCCs

Figure 4 shows the time traces for a set of repeat discharges, which have a plasma current of 400 kA and flat top currents in the poloidal field coils of $I_{P4} = - 59$ kAt and $I_{P5} = -107$ kAt (giving $I_{P5}/I_{P4} = 1.8$). When the plasma had evolved to steady conditions, i.e. constant density and regular sawteeth, the current in one of the pairs of error field coils was ramped up until an error field mode was formed (see Figure 4b and c). These modes are formed locked and have no rotating m = 2, n = 1 signature. The locked mode is detected by an array of saddle coils which measure the radial magnetic field at the vacuum vessel on the outer mid-plane of the machine. On formation of a locked mode a growing n = 1 perturbation is seen on the saddle coils as shown in Figure 4d. This is accompanied by a drop in the plasma density (Figure 4d) and in this case by a rapid termination of the shot. The shot was repeated with a different phase of the applied field (i.e. either the other sign of the current or the other pair of coils). An example of the currents that were required to



trigger the onset of a locked mode for the four different phases of the applied error field is shown in Figure 4. The current shown has been corrected for the penetration through the vessel wall, which has a time constant of 31 ms and represents the current that would produce the effective field at the edge of the plasma. As can be seen the plasma locks at a different value of the current for each of the four phases.

The results shown in Figure 4 where obtained for a plasma with a high ratio of the $I_{P5}/I_{P4}=1.8$. Figure 5 is a shot with a low value of $I_{P4}/I_{P5} = 0.25$ (i.e. $I_{P4}$ = -172 kAt, $I_{P5}$ = -43.5 kAt). Again as the current in the EFCCs is ramped up until the onset of a locked mode, in this case this leads to a density pump out but there is no rapid termination of the plasma. Figure 6a and b show the current in the EFCC coils at the onset of the locked mode as a function of the phase of the applied field. In both cases the points lie on approximately on a circle. The radius of the circle is much smaller in Figure 6b due to the lower density of the plasma in this shot, which is consistent with the fact that the locked mode threshold scales as $\sim n_e^{1.1}$[7]. The applied error field, is directly proportional to the current in the error field coils, so the radius of this circle is proportional to the total error field required to form the locked mode, and the centre of the circle is identified as giving the value of the EFCC currents that will best minimize the n=1 component of the intrinsic error field for that specific plasma equilibrium. In the both cases the optimum correction current in EFCC_5_11 ($I_{EFCC511}$) is negative, while the current in EFCC_2_8 ($I_{EFCC28}$) changes from negative for the scenario with $I_{P5}/I_{P4}$= 0.25 to positive for the plasma with $I_{P5}/I_{P4}$= 1.8. The empirically derived optimum error field correction for the shot with $I_{P5}/I_{P4}$= 1.8 has $I_{EFCC28}$ = -1.7 kAt and $I_{EFCC511}$ = -2.8 kAt while for the shot with $I_{P5}/I_{P4}$= 0.25 the values are $I_{EFCC28}$ = 0.8 kAt and $I_{EFCC511}$ = -1.6 kAt.



This technique has been repeated for all the standard plasma scenarios used on MAST and the optimum error field correction current in each case has been determined from the centre of the circle produced. Figure 7 shows a plot of the EFCC coil current expressed as a fraction of the current in the P4 coil versus the ratio of the P5 to P4 currents. As can be seen for both sets of EFCCs there is a linear dependence between the required correction and the ratio of the current in the poloidal field coils, which suggests that the correction does have a strong dependence on the currents in these coils. This means that the required EFCC correction current can be determined by a fit to the form $I_{EFCC} = \alpha I_{P4} + \beta I_{P5}$ (i.e. $I_{EFCC}/I_{P4} = \beta I_{P5}/I_{P4} + \alpha$). The fits lead to the straight lines shown in Figure 7 and results in $\alpha_{28}$ = -0.010±0.001, $\beta_{28}$ = 0.021±0.001, $\alpha_{511}$= -0.001±0.001 and $\beta_{511}$ = 0.028±0.002 (the currents used in P4 and P5 are normally negative). The dotted lines shown in Figure 7 represent the ±1σ uncertainty in the fits. A lot of the discharges studied have $I_{P5}/I_{P4}$~0.7 and there is some scatter in the empirically determined correction for these shots. This scatter most likely results from the uncertainty in the individual measurements as well as indicating that there is a possible contribution to the intrinsic error field from other sources not considered.

The EFCC correction current as a function of poloidal coil current has been coded into the MAST plasma control system, to enable a real time correction of the error field based on the currents in the P4 and P5 coils. The typical waveforms produced are shown in Figure 8 and are compared to the previous static error field corrections, which were based empirically on the plasma current and toroidal field of the shot in question [7]. At flat top the new correction has a similar size to the previously derived correction and as such



produces a similar effect on the plasma, however, this new correction is based on what is thought to be the source of the intrinsic error field rather than a scaling alone.

To demonstrate the effectiveness of this new error field correction algorithm, over a shot without error field correction, a discharge has been developed where the density was gradually reduced from a flat top value. This shot has $I_P$ = 600 kA, $q_{95}$ = 7.0, $I_{P4}$ = -145 kAt and $I_{P5}$ = -90 kAt. Figure 9 shows that in the case where there was no error field correction, a locked mode starts to grow from 170 ms. There is a sudden drop in density (Figure 9a), the sawteeth disappear from the soft X-ray trace (Figure 9c) and the discharge disrupts at 270 ms. With error field correction, the density in the discharge continues to drop and the sawteeth continue until a locked mode is finally observed at 285 ms, when the density has decreased by 45 % (from $1.21 \times 10^{19}$ to $0.55 \times 10^{19}$ m$^{-3}$) relative to the shot without error field correction.

## 4.2 Determination of the intrinsic n=1 error field correction using the ELM coils

A correction of the n=1 intrinsic error field can also be produced using the internal ELM coils[18]. The 6 upper ELM coils are used with 6 of the lower ELM coils. To produce an n=1 field, in the upper row of coils a current of opposite sign ($I_{ELM}$) is applied to two coils separated by 180°. The neighbouring two coils carry a current of $I_{ELM}/\sqrt{2}$. For the bottom row a similar configuration of full and $1/\sqrt{2}$ current is used but the location of the coil with full current is chosen such that the pitch angle of the applied field lines up with the pitch angle of the plasma equilibrium field at the q=2 surface. Figure 10 shows an example of this where the perturbation has been aligned with the q=2 surface for a shot with $I_P$ =



600 kA. The maximum positive radial field ($B_r$) in the upper row of coils is located at sector 9, corresponding to $\phi = 3.4$ radians.

The current that can be applied to the ELM coils is not sufficient to routinely produce a locked mode. Therefore the discharge used in the previous section to verify the error field correction from the EFCCs is the one used in which the density is gradually reduced from a flat top value at constant applied field. Figure 11 shows a matched pair of shots, with and without applied error field correction from the ELM coils. For both shots there was no current applied in the external EFCCs. In the case where no error field correction was applied, soon after the density is reduced a locked mode starts to grow from ~170 ms (determined from the change in the radial magnetic field measured by the saddle coils, Figure 11c). With error field correction using $I_{ELM} = 4.0$ kAt in the ELM coils, the density of the discharge continues to drop until a locked mode is formed at ~240ms, by which time the line averaged density has dropped by 60%.

This discharge has been repeated using different values of $I_{ELM}$ and the minimum density achieved before the onset of a locked mode established. Figure 12a shows a plot of the minimum or critical density ($n_e^{crit}$) achieved before the onset of a locked mode, as determined from the deviation of signals in the saddle coils, as a function of the current in the upper ELM coil located at sector 9. Note negative currents are equivalent to rotating the pattern of current by $\pi$ radians. Positive values of $I_{ELM}$ clearly reduce the minimum density that can be obtained before the onset of a locked mode, however, due to the limit on the maximum current ($I_{ELM} = 5.6$ kAT) it is not clear whether the optimum correction has been obtained. Negative values of the current clearly make the locked mode occur at higher



density. To obtain these measurements the shots had to be repeated at higher initial density than that shown in Figure 11, otherwise the locked mode formed immediately i.e. before the density ramp down started. Shown in Figure 12a as the horizontal dashed line is the minimum density achieved using the optimum error field correction form the external EFCCs. Hence a similar level of correction, in terms of minimum density, can be achieved using either the external or internal coil sets.

By shifting the pattern of currents toroidally in the upper and lower rows of the RMP coils (i.e. moving the location of the coil with maximum $B_r$ toroidally by one coil at a time) the phase of the applied field relative to the intrinsic error field can be rotated in steps of 60°. The minimum density obtained before the onset of the locked mode has then been determined as a function of this angle for a range of $I_{ELM}$, see Figure 11b. Full scans have been completed for $I_{ELM}$ = 2,0 and 4.0 kAt, while for $I_{ELM}$=5.6 kAT only point around the minimum could be performed due to the ease of locking at other phases.

The total radial field at a particular location can be expressed as

$$b_{res}^r = \sqrt{b_{intr}^2 + b_{corr}^2 + 2b_{intr}b_{corr}\sin(\varphi - \varphi_0)}$$ where $b_{intr}$ is the intrinsic component $b_{corr}$ is the applied correction field and $\varphi_0$ is the angle between the intrinsic and applied fields, and that the minimum densities are related to these fields. Then assuming that this radial field is linearly related to the critical density at which the locked mode occurs the data in Figure 12b can be fitted to the form

$$n_e^{crit} = \sqrt{n_{intr}^2 + n_{corr}^2 + 2n_{intr}n_{corr}\sin(\varphi - \varphi_0)}$$

where $n_{intr}$ is the locking density with no correction and $n_{corr}$ would be the locking density if there were no intrinsic error field and only the correction field was applied. This fits



results in a good description of the data. The results of the fit will be compared to the modelled correction in section 5.1.

## 5. *Comparison with vacuum and plasma response modelling*

In order to investigate how the empirically derived error field corrections relate to the measured error fields due to the non-uniformities in the P4 and P5 poloidal coils vacuum magnetic modelling has been performed using the ERGOS code [19]. The vacuum response may not be an appropriate measure of the external field at the q=2 surface, due to the effect of screening and because it is known that the plasma response is important near to a stability limit [20]. While the plasmas studied in this paper are far from stability limits, previous studies on DIII-D and NSTX using the IPEC code [21] have shown that the plasma response changes the assumed superposition of the fields even for a plasma far from a stability limit. These effects have been investigated using the MARS-F code [22], which is a linear single fluid resistive MHD code that combines the plasma response with the vacuum perturbations, including screening effects due to toroidal rotation.

### 5.1 Vacuum field modelling using the ERGOS code

The $B_r$, $B_z$ and $B_\phi$ components of the magnetic fields have been calculated for the intrinsic error fields, the error field correction coils and ELM coils on a 3D grid. For each poloidal field coil the fields are calculated using the distortions described in Table 1. The fields due the EFCCs and the ELM coils are calculated using a realistic description of the coils. The fields are then added together according to the current in each coil and are



combined with the plasma equilibrium field. The n=0 component of the intrinsic error fields are subtracted since they are already included in the magnetic equilibrium.

To demonstrate the size and structure of the intrinsic error field the radial component of the magnetic field at the location of the q=2 surface at LFS mid-plane (r= 1.258 m, z=0 m) for the plasma described in the previous section (which has $I_{P4}$ = -140 and $I_{P5}$ = -90 kAt) is shown as a function of toroidal angle in Figure 13a (solid line). There is a ~ 10 G peak to peak size of the error field which has a complex toroidal structure. A fast Fourier transform (FFT) of the signal is shown in Figure 13b that reveals that the field is composed of not only an n=1 component but also a sizeable n=2 component.

The dotted curve in Figure 13a shows the field due to the EFCCs which has been calculated using the currents found empirically to optimise the error field correction (i.e. $I_{EFCC28}$ = -0.46 kAt, $I_{EFCC511}$ = -2.2 kAt). The FFT of this field (Figure 13b) shows that it is dominated by an n=1 contributions with a small n=3 sideband. There is no even-n component due to the layout of the coils. Finally the dashed curve is the sum of the intrinsic error field and EFCC fields. While at some toroidal angles the radial field is reduced at others it is increased. The FFT of the combined field shows that although the n=1 component is reduced, it is not zero and also that now the dominant component has a toroidal mode number n = 2.

While looking at a single location can help to demonstrate the problem of correcting the intrinsic error field, what is generally considered to be the most important in parameter in the onset of locked modes is the resonant component of the magnetic field normal to the q = 2 surface averaged over the flux surface. The normalised component of the perturbed field perpendicular to equilibrium flux surfaces is given by



$b^1{}_{mn} \equiv \left[ \left( \vec{B} \cdot \vec{\nabla} \, \psi^{\frac{1}{2}}_{pol} \right) \Big/ \left( \vec{B} \cdot \vec{\nabla} \, \varphi \right) \right]_{mn}$ where m is the poloidal mode number, n the toroidal mode

number, $\vec{B}$ is the total field vector, $\Psi^{\frac{1}{2}}_{pol}$ the square root of the poloidal flux and $\varphi$ is the

toroidal angle [24]. This can then be normalised to the toroidal field to give the resonant

field component on each rational surface ($b^r{}_{res}$). Figure 14 shows the values of $b^r{}_{res}$ for the

n=1 toroidal field component as a function of normalised radius for the intrinsic (circle) and

empirically determined error field correction from the EFCCs, together with the location of

the q=2 surface. Although the addition of the field from the EFCCs reduces $b^r{}_{res}$ at the q=2

surface it is not zero. As shown in Figure 14 as the open squares, the value of $b^r{}_{res}$ for the

n=1 toroidal field component can be made equal to 0 at the q=2 surface by adjusting the

values of the currents in the EFCC coils such that $I_{EFCC28}$ = -0.154 kAt and $I_{EFCC511}$ = -0.865

kAt.

The empirically derived field is a factor of ~3 larger than that determined from

vacuum modeling but the direction of the correction is similar (differing by 10 degrees).

The modeling has been performed for all the discharges for which the empirically

determined optimum error field correction has been established and a similar trend is

observed. Figure 15 shows a plot of the optimum error field correction current derived

from minimizing the value of $b^r{}_{res}$ at the q=2 surface using vacuum modeling versus the

empirically determined correction for these discharges. In the case of $I_{EFCC28}$ the data is

consistent with a linear fit going through the origin with a slope of 0.35±0.02. In the case

of $I_{EFCC511}$ constraining the fit to go through the origin gives a slope of 0.34±0.03. However

there is a clear outlier at $I_{EFCC511}{}^{emp}$ = -1.67 kAt, $I_{EFCC511}{}^{pred}$ = -1.08 kAt. Removing this

outlier and performing a free fit give the dashed line in Figure 15 which has a slope of



0.37±0.04 and an offset of 0.018±0.03. The difference between the predicted value and the empirically derived values and the existence of the outlier could be due to a poor understanding of intrinsic error fields and/or a need to take into account plasma response, which will be discussed in the next subsection.

Similar vacuum modelling calculations have been performed for the shots in which the error field correction is applied using the ELM coils. Figure 16 shows the values of $b^r_{res}$ for the n=1 toroidal field component as a function of normalised radius for the intrinsic (circle) and empirically determined error field correction from the ELM coils, together with the location of the q=2 surface, for the shot discussed in Figure 11. Similar to what is observed with the EFCCs, the addition of the field from the empirically derived ELM coil current ($I_{ELM}$=4 kAt) s reduces $b^r_{res}$ at the q=2 surface but it is not zero. In the case of the ELM coils because of the fixed angle of the applied perturbation, it is not always possible to find a value of the coil current that produces reduces $b^r_{res}$ to 0. In this configuration of the coils a value of $I_{ELM}$ = 2.3 kAt is found to minimise the value of $b^r_{res}$ at the q=2 surface. Similar to what was observed with the EFCCs the value of the optimized coil current predicted by vacuum modeling is much smaller than the value found by the locked mode studies.

Calculations have been performed of $b^r_{res}$ as a function of the phase angle and size of the applied correction from the ELM coils. Figure 17 shows the that the optimum phase angle, defined as the minimum $b^r_{res}$ is very close to the one that can be applied experimentally due to the discrete nature of the coils. Superimposed on Figure 17 as the solid line is the results of the fit to the critical density for the onset of the locked mode determined from the fit to the data in Figure 12b. Similar to what was observed with the



EFCCs, although the magnitude of the predicted and empirically determined correction differ by more than a factor of 2 the difference in the direction of the perturbation is similar.

## 5.2 Plasma response modelling

The effect that the plasma response has on the error field correction from the EFCCs has been investigate using the MARS-F code, which is a linear single fluid resistive MHD code that combines the plasma response with the vacuum perturbations, including screening effects due to toroidal rotation [22]. The calculations use the experimental profiles of density, temperature and toroidal rotation as input and realistic values of resistivity, characterised by the Lundquist number (S) which varies from $\sim 10^8$ in the core to $\sim 10^6$ in the pedestal region (the radial profile of the resistivity is assumed proportional to $T_e^{-3/2}$). The intrinsic and error field correction coils have been represented in the model as an equivalent surface current [23]. The advantage of this approach is that the error fields can be included in the code without an exact knowledge of their sources. All that is required is that the error field are specified as a normal field at a surface just outside the plasma.

Modelling has been performed for a range of MAST plasmas in order to investigate the effect of the plasma response and several correction criteria have been investigated [23]. While no single criteria are in agreement with all the empirically derived corrections, the two that are in best overall agreement with the empirically derived corrections are minimising the n=1 m=2 resonant field component at the q=2 after the plasma response has been taken into account and minimising the overall jxB torque on the plasma.



Figure 18a shows the current in the EFCCs for several criteria for shot 26467, which has $I_{P4}$ = -56 and $I_{P5}$ = -105 kAt. As discussed above the empirically determined correction is ~ 3 times larger than the coil currents required to minimise the n = 1 resonant field component at the q=2 surface in the vacuum approximation. However, when the plasma response is included, in this case, the calculated coil currents are found to be in much better agreement with the empirically derived values. At first this may appear strange as it may have been expected that the plasma response would screen the intrinsic field and the applied correction field at a similar level. However, the difference in the poloidal spectrum of the two fields means that, after the plasma response is taken into account, the resulting field at the q=2 surface is modified.

For shot 26467 the coil currents required to minimise the total jxB torque on the plasma are not in good agreement with the empirically derived corrections but this is not always the case. Figure 18b shows the coil currents for the different criteria for shot 26051 which has $I_{P4}$ = -100 and $I_{P5}$ = -76 kAt. In this case the empirically determined error field correction is in better agreement with the jxB criteria than the minimisation of the resonant field component. Whilst the corrections calculated taking into account the plasma response are always better than those calculated in the vacuum approximations as shown in Figure 18c for shot 10738, which has $I_{P4}$ = -140 and $I_{P5}$ = -87 kAt the agreement is not always good. In this example the empirically determined correction is half way between the two corrections, which take into account the plasma response.

One thing that is noticeable in all the cases shown in Figure 18 is that while the amplitude of the correction currents is different for the various criteria the direction of the applied field is similar. The difference between the empirically determined correction and



the various criteria in terms of amplitude and phase are shown in Figure 19a and b for all the discharges for which MARS-F simulations have been performed. Overall the best criteria is the one based on minimising the n = 1 resonant field component on the q=2 surface taking into account the plasma response.

### 6. Summary and discussion of future improvements

Previous studies have indicated that the dominant intrinsic error fields in MAST result from distortions in the poloidal field coils. The radial and vertical field generated by the P4 and P5 poloidal field coils have been measured as a function of toroidal angle using a Hall probe. The resulting distributions have been fitted assuming that they are due to a set of distortions to the shape of these coils. This parameterisation has then been used to determine a 3D map of the intrinsic error fields in MAST that can be used in modelling for comparison with experimental measurements.

The correction of the n=1 component of the intrinsic error field can be performed either using the external error field correction coils or the internal ELM control coils. Two methods have been used on MAST to determine the onset criterion for locked modes and hence determine empirically the optimum error field correction: Either ramping the applied field at constant density or using a decreasing density at fixed applied field. In the case of the ELM coils, due to the current available only the second method could be used. The empirically derived correction has been determined for a wide variety of MAST shots. The empirically derived currents required in the EFCCs scale with the currents in the P4 and P5 coils suggesting that these poloidal field coils are the dominate source of the n=1 error field in MAST. If the empirically derived currents are applied the discharges can be run at lower



density without the onset of a locked mode, which considerably increases the operation space on MAST.

The intrinsic error field on MAST has been parameterised in terms of the distortions to the P4 and P5 poloidal field coils, which has been used as input to both vacuum and plasma response modelling. The onset of locked modes is thought to be due to the formation of islands on the q=2 surface, which is related to the size of the n=1 resonant radial field at this location. Therefore the current in the EFCCs required to minimise this value has been calculated. In the vacuum approximation the current required is approximately a factor of 3 times smaller than that found empirically. When the plasma response is included there is better agreement between the empirical and predicted values but the agreement is not perfect. Other criteria have been considered, one of the most promising is the minimisation of the jxB torque on the plasma, but while this can explain some of the observed differences it can not explain everything. It is likely a combination of several criteria is required, including a study of the non-linear coupling of different processes. These studies could be improved if measurements were made of the response of the plasma to the applied field using specially optimised magnetics diagnostics. Such diagnostics are currently being designed for installation in the ongoing upgrades to MAST.

Even if the n = 1 component can be corrected there is still a large residual n = 2 component. Experimental evidence for an n=2 intrinsic error field has been obtained during ELM control experiments on MAST using an n = 2 RMP field. In these experiments it has been found that depending on the phase of the applied perturbation either ELM mitigation or a locked mode results. The correction of the n = 2 component of the error field will be the subject of future work.



## Acknowledgement

This work was part-funded by the RCUK Energy Programme [grant number EP/I501045] and the European Communities under the contract of Association between EURATOM and CCFE. To obtain further information on the data and models underlying this paper please contact PublicationsManager@ccfe.ac.uk. The views and opinions expressed herein do not necessarily reflect those of the European Commission

**Tables**

**Table 1** Results from the fits to the measured radial and vertical fields for the P4 and P5 coils (shown in Figure 2 and Figure 3)in terms of shifts, tilts and coil deformations (as described in the text).

| PF Coil | P4U | P4L | P5U | P5L |
|---|---|---|---|---|
| Current (kA) | 3.746 | 3.692 | 3.595 | 3.661 |
| X shift – x (mm) | 1.269 | 3.183 | -3.276 | 0.756 |
| Y shift – y (mm) | 0.342 | -2.613 | 1.453 | -2.245 |
| Z-Shift – $\delta z_0$ (mm) | 2.941 | 6.638 | 0.562 | 13.07 |
| Angle of tilt – $\lambda$ (degrees) | 0.1189 | -0.0865 | -0.0338 | 0.1192 |
| Axis of tilt – $\theta_t$ (degrees) | 15.2 | -89.0 | 31.7 | 58.6 |
| Ellipticity – $\varepsilon$ (mm) | 1.7667 | 3.752 | -2.839 | 3.1913 |
| Axis of ellipse – $\theta_\varepsilon$ (degrees) | -1.4 | 36.0 | 28.2 | 15.8 |
| Banana deformation – $\beta$ (mm) | 3.772 | -2.183 | -4.916 | -3.823 |
| Axis of banana – $\theta_\beta$ | -23.0 | -3.0 | -18.8 | -101.3 |
| Wobble deformation $\omega$(mm) | 0.702 | 1.727 | -0.658 | -1.469 |
| Axis of wobble – $\theta_w$ | -58.6 | -106.1 | -191.7 | -38.8 |



**Figures**

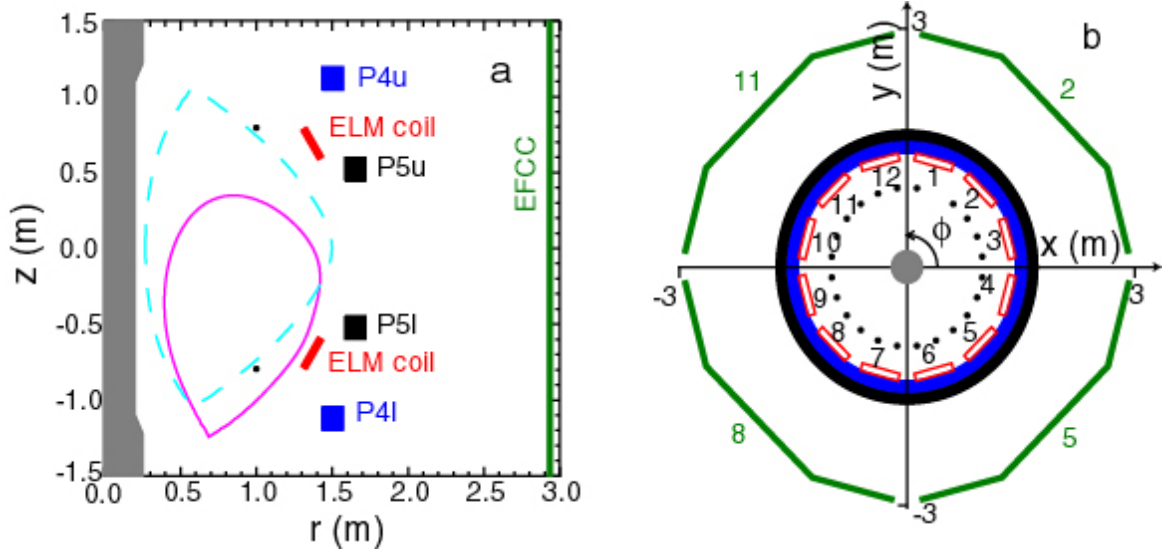

**Figure 1** Location of the poloidal field coils (P4 and P5), the external error field correction coils (EFCC) and internal ELM coils in a) the poloidal cross section and b) the plan view on which are shown the definition of the sector numbers and the toroidal angle ($\phi$). The dots in a) and b) show the location of the Hall probe used to measure the field due to the P4 and P5.



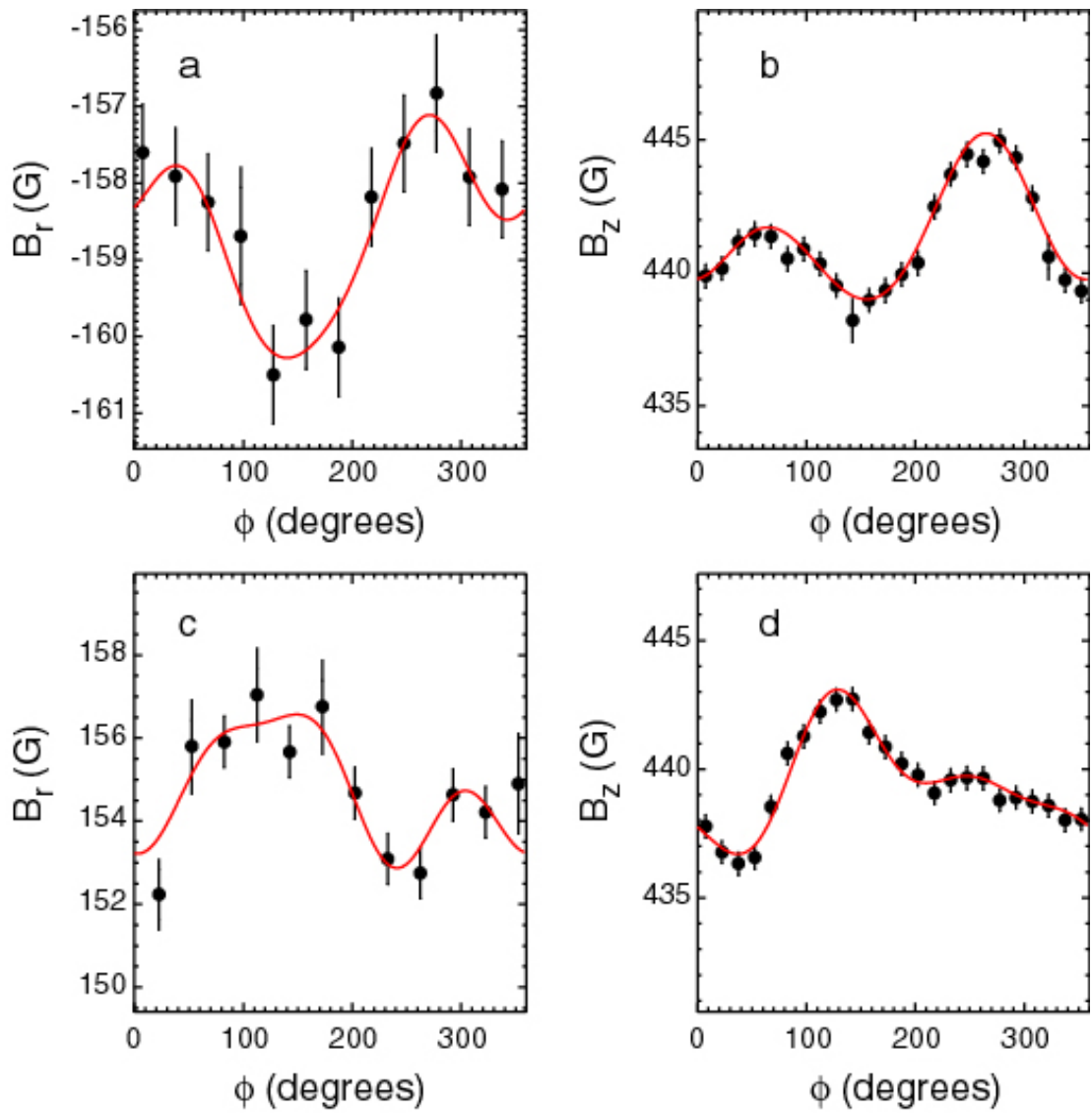

**Figure 2** Measurements of the radial (B$_r$) and vertical (B$_z$) magnetic field as a function of toroidal angle ($\phi$) due to a flat top current of -3.5 kA (-80.5 kAt) in the a),b) upper and c),d) lower P4 coils.



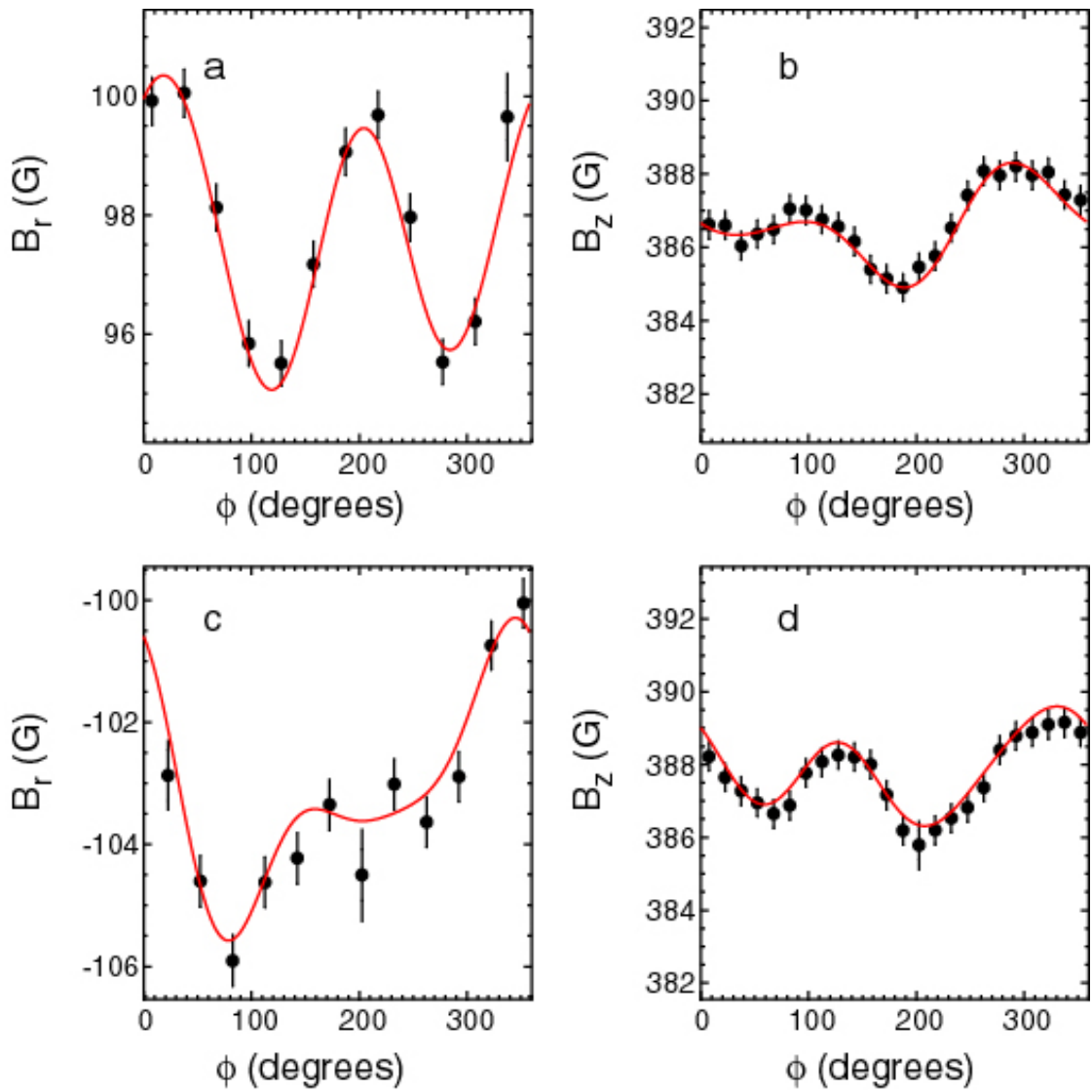

**Figure 3** Measurements of the radial ($B_r$) and vertical ($B_z$) magnetic field as a function of toroidal angle ($\phi$) due to a flat top current of -3.5 kA (-80.5 kAt) in the a),b) upper and c),d) lower P5 coils.



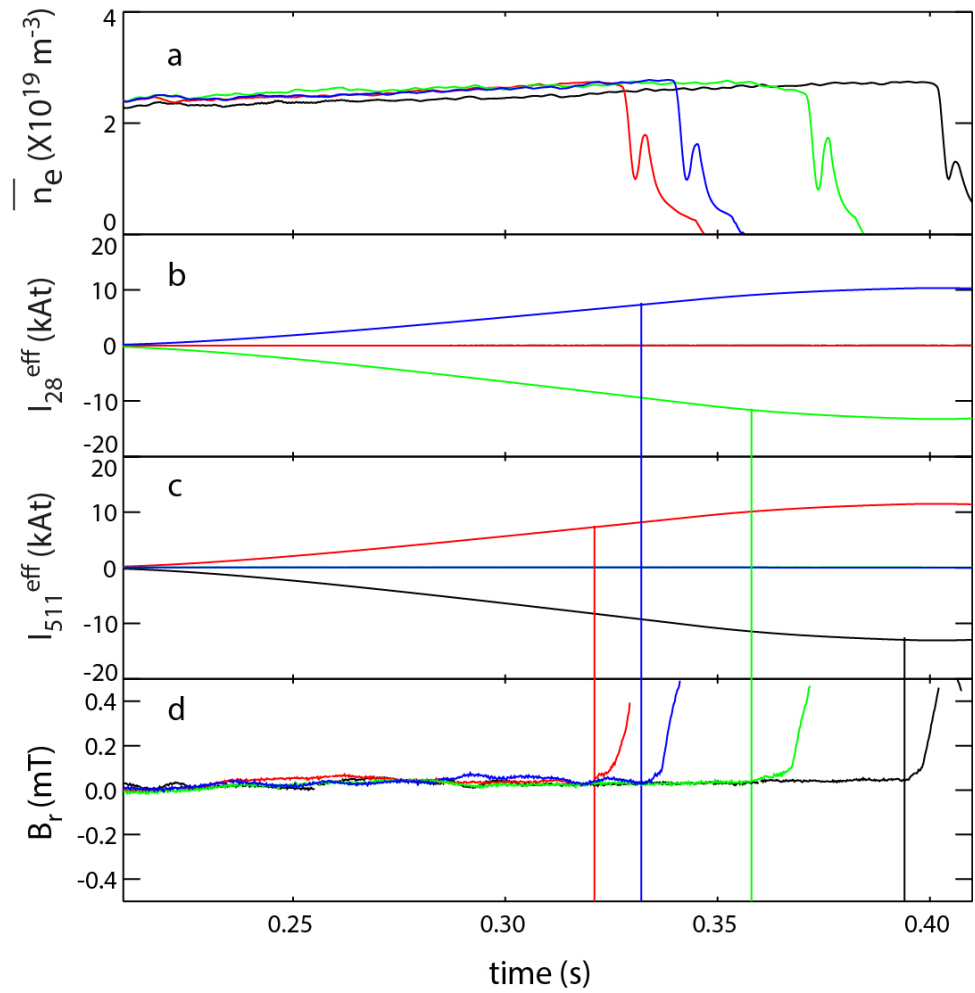

**Figure 4** Time traces for a series of shots with $I_P$=400 kA, with $I_{P5}/I_{P4}$=1.8 of a) line average density ($\bar{n}_e$), the effective current in the EFCC pair b) 2-8 ($I_{28}^{eff}$) and c) 5-11 ($I_{511}^{eff}$) and d) the radial field component of the magnetic field determined from an array of saddle coils.



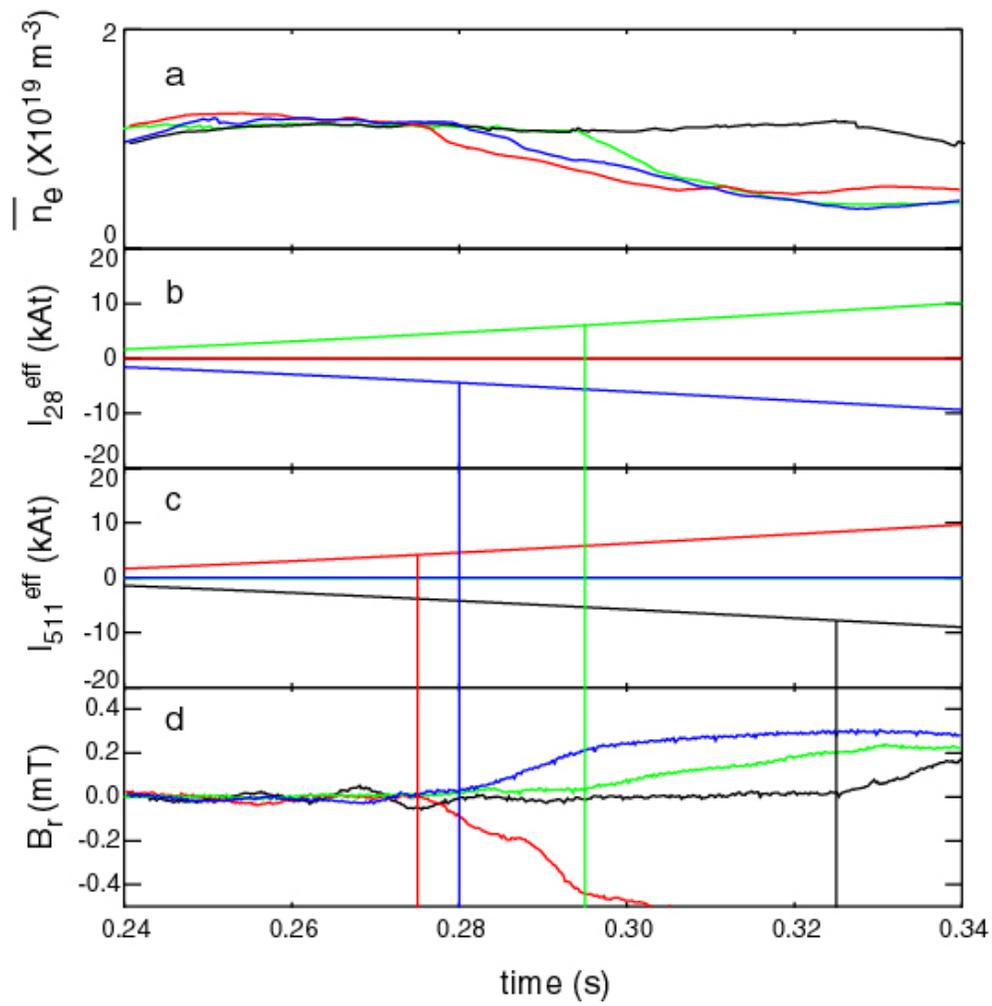

**Figure 5** Time traces for a series of shots with $I_P$=400 kA, with $I_{P5}/I_{P4}$=0.25 of a) line average density ($\bar{n}_e$), the effective current in the EFCC pair b) 2-8 ($I_{28}^{eff}$) and c) 5-11 ($I_{511}^{eff}$) and d) the radial field component of the magnetic field determined from an array of saddle coils.



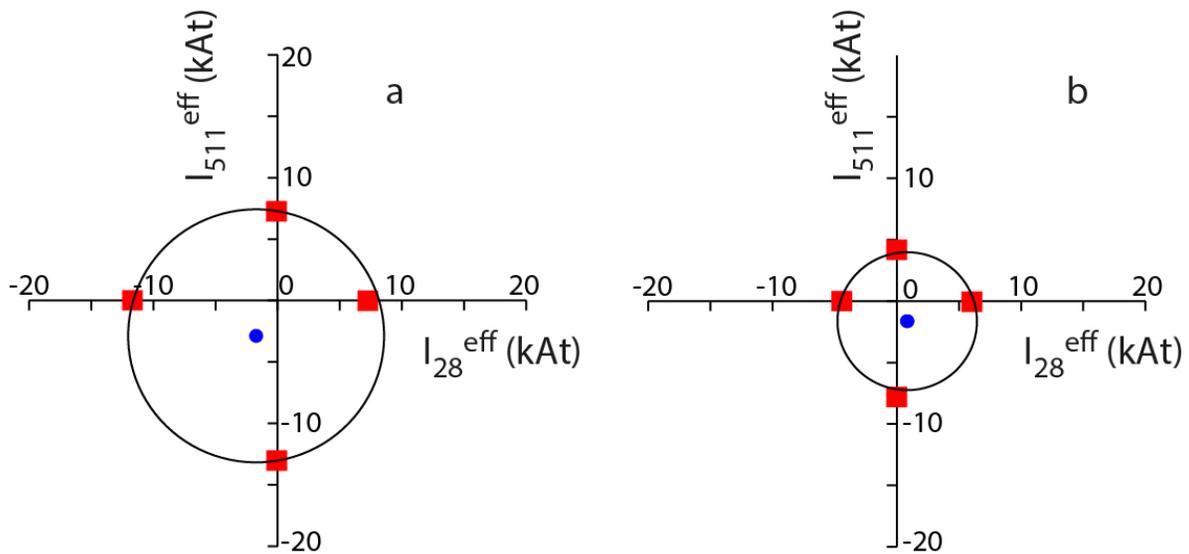

**Figure 6** The error field coil currents required to form a locked mode for an equilibrium with Ip = 400 kA with a) $I_{P5}/I_{P4}$= 1.8 and b) $I_{P5}/I_{P4}$=0.25.

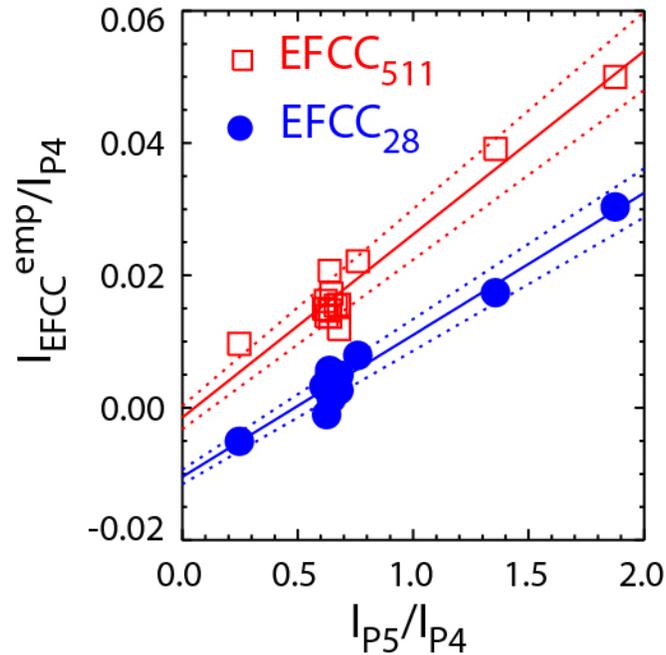

**Figure 7** The empirically derived error field correction current ($I_{EFCC}$) expressed as a fraction of the current in the P4 coils ($I_{P4}$) versus the ratio of currents in P5 over P4 ($I_{P5}/I_{P4}$)for the 2-8 (circle) and 5-11 (square) EFCC pairs. The solid lines show the results of a linear fit to the data and the dotted lines the ±1σ contours.



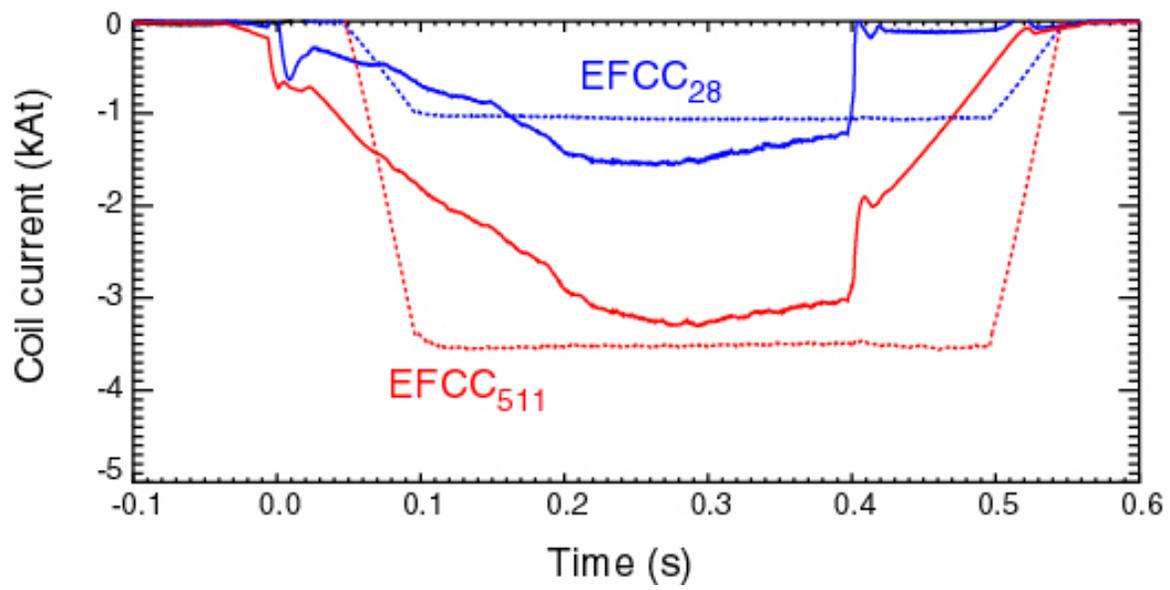

**Figure 8** Example of time traces of the current waveforms in the EFCC for the old (dashed) and new (solid) correction waveforms.



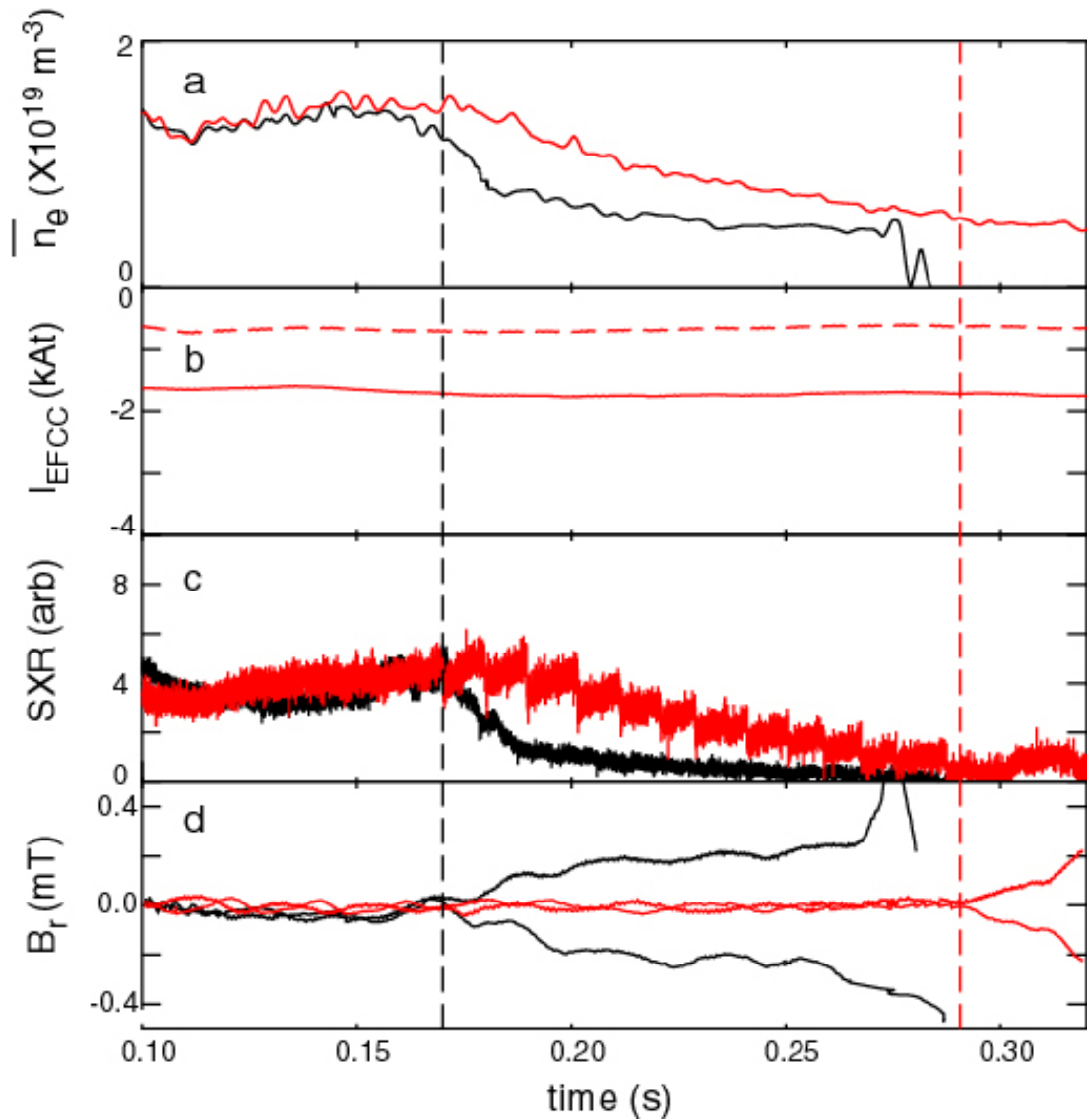

**Figure 9** Time traces of a) line average density ($\bar{n}_e$), b) the effective current in the 2-8 (dashed) and 5-11 (solid) EFCC pairs c) the soft X-ray signal from a core viewing channel and d) the radial field component of the magnetic field determined from an array of saddle coils for a shot without (black) and with (red) error field correction.



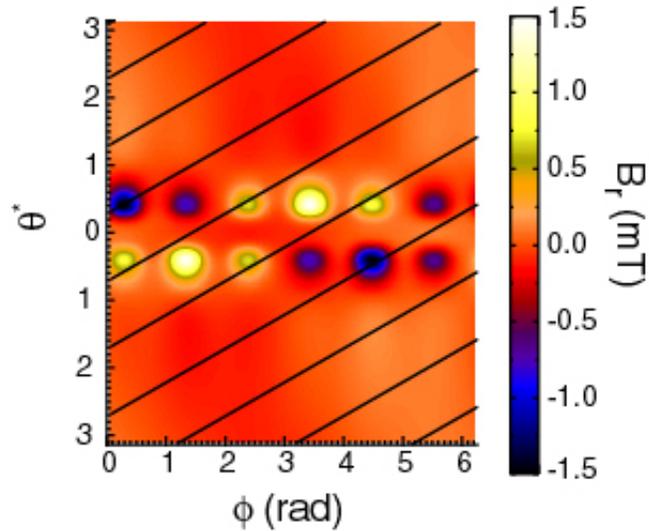

**Figure 10** Plot of the component of the magnetic perturbation (vacuum approximation) perpendicular to the equilibrium flux surfaces as a function of toroidal angle ($\phi$) and the poloidal angle in a straight field line co-ordinate system ($\theta*$) on the q=2 flux surface for discharges with $I_P$ = 600 kA for a configurations designed to reduce the n=1 component of the intrinsic error field. The solid lines show the equilibrium field lines at the q=2 surface.



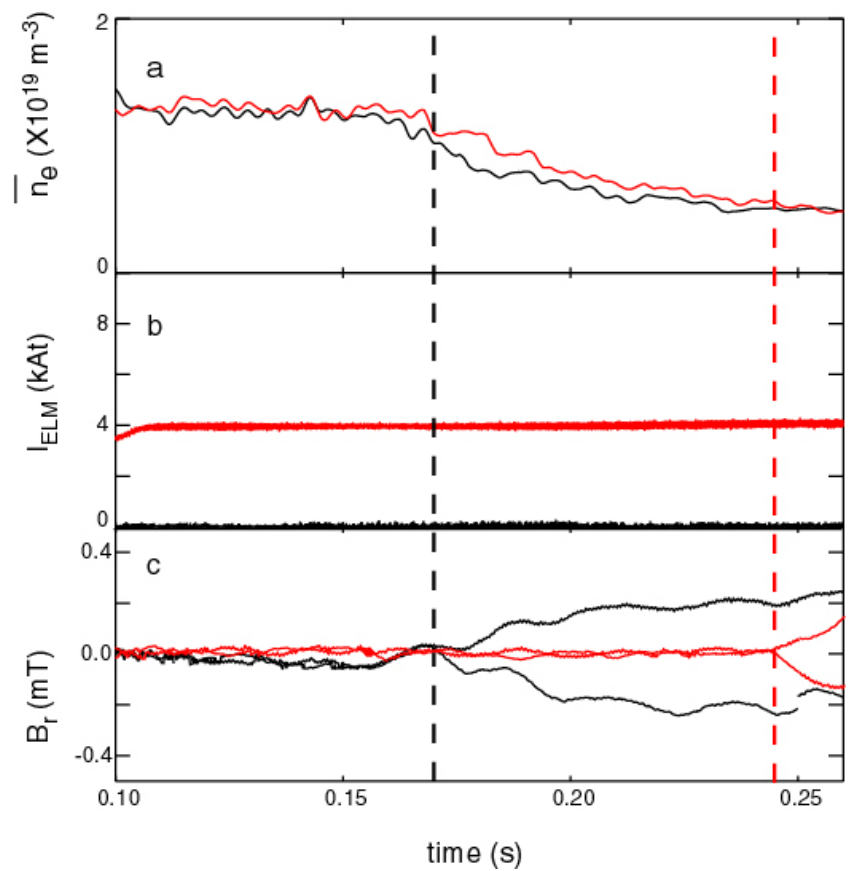

**Figure 11** Time traces of a) line average density ($\bar{n}_e$), b) the current in the ELM coils ($I_{ELM}$) and c) the radial field component of the magnetic field determined from an array of saddle coils.



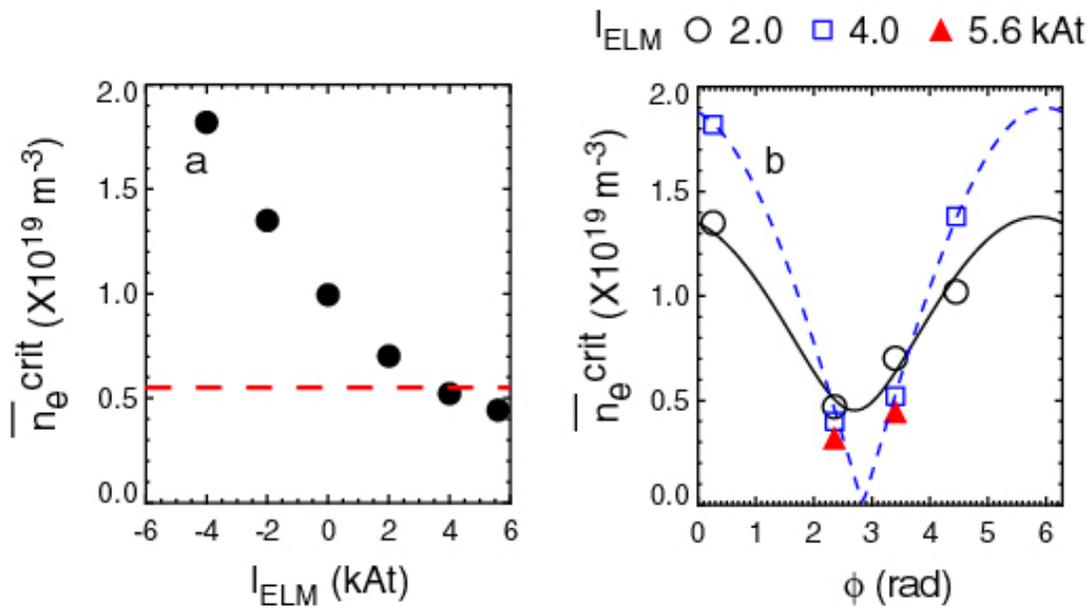

**Figure 12** a) the density at the onset of the locked mode ($n_e^{crit}$) versus current in the ELM coils ($I_{ELM}$). The dashed line shows the critical density achieved using the optimum error filed correction form the external EFCCs. b) $n_e^{crit}$ as a function of the toroidal angle of the applied correction field for currents in the coils $I_{ELM}$ = 2.0 (open circle), 4.0 (square) and 5.6 (triangle) kAt. The curves are a result of a fit to the applied and intrinsic n=1 fields.



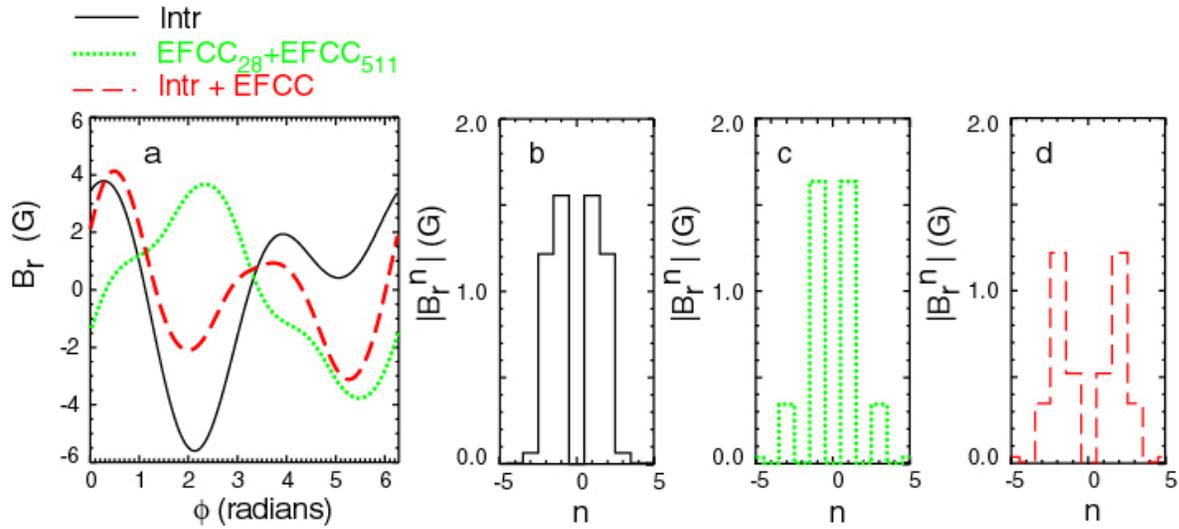

**Figure 13** a) The radial field at the location of the q=2 surface at the low field side midplane, calculated using vacuum modelling due to the intrinsic error field only (solid), EFCCs only (dotted) and corrected field (dashed). The toroidal components of the radial field for b) the intrinsic error field, c) the EFCCs and d) the corrected field.

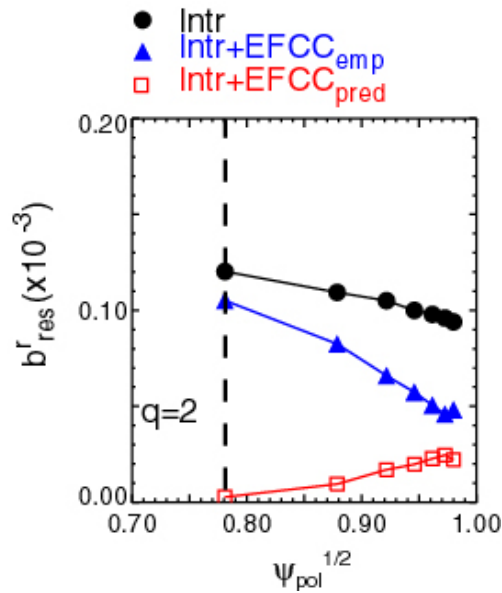

**Figure 14** Calculations in the vacuum approximation of the normalised resonant component of the applied field ($b^r_{res}$) for the n=1 components of the intrinsic only (circle), intrinsic plus empirically determined error field correction from the EFCCs (triangle) and the intrinsic plus optimum predicted correction (square). The vertical dashed line shows the location of the q=2 surface.



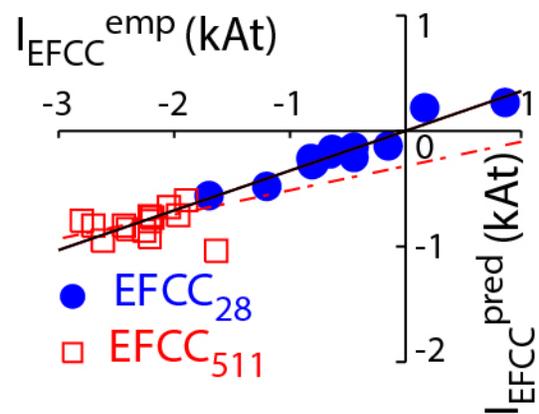

**Figure 15** The empirically derived error field correction from the EFCCs versus the optimum derived from vacuum modelling for the 2-8 (circle) and 5-11 (square) EFCC pairs. The lines represent a linear fit to the data.



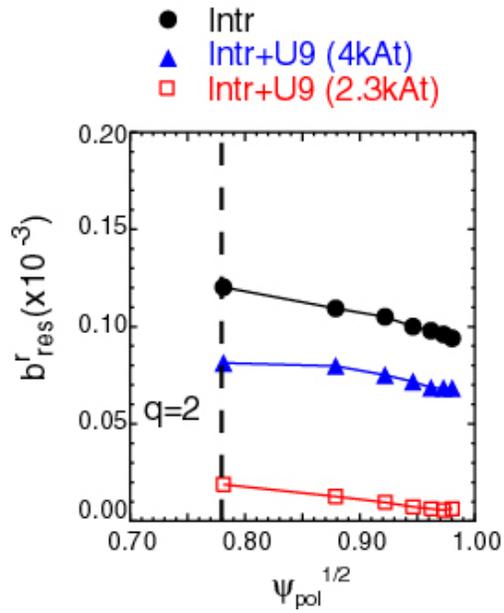

**Figure 16** Calculations in the vacuum approximation of the normalised resonant component of the applied field ($b^r_{res}$) for the intrinsic only (circle), intrinsic plus empirically determined error field correction from the ELM coils (triangle) and the intrinsic plus optimum predicted correction (square). The vertical dashed line shows the location of the q=2 surface.



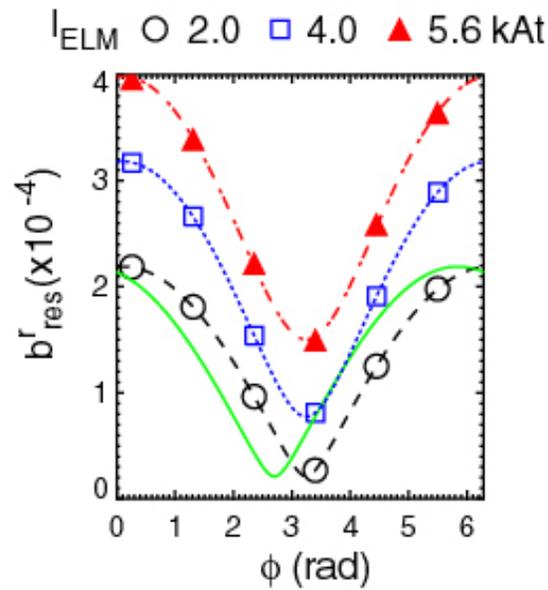

**Figure 17** The effect of changing the amplitude and phase angle of the correction field from the ELM coils on the normalised resonant component of the applied field ($b^r_{res}$). The dashed curves are a fit to the intrinsic and applied field described in the text. The solid curve is the results of the fit to the empirically critical density.

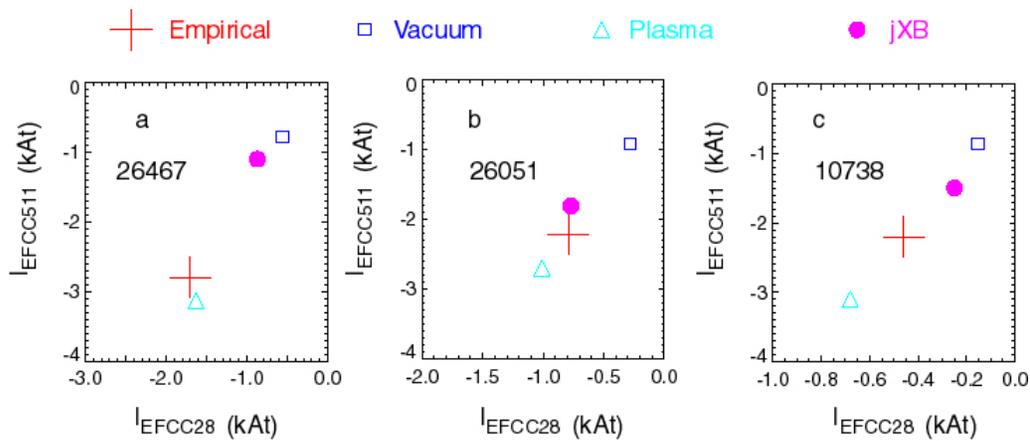

**Figure 18** Coil in the EFCC correction coils found to empirically correct the intrinsic error field correction (cross), to minimise the n=1 m=2 radial field component at the q=2 surface using the vacuum approximation (square) or taking into account the plasma response (triangle) and minimising the jxB torque on the plasma (circle) for shots a) 26467, b) 26051 and c) 10738.



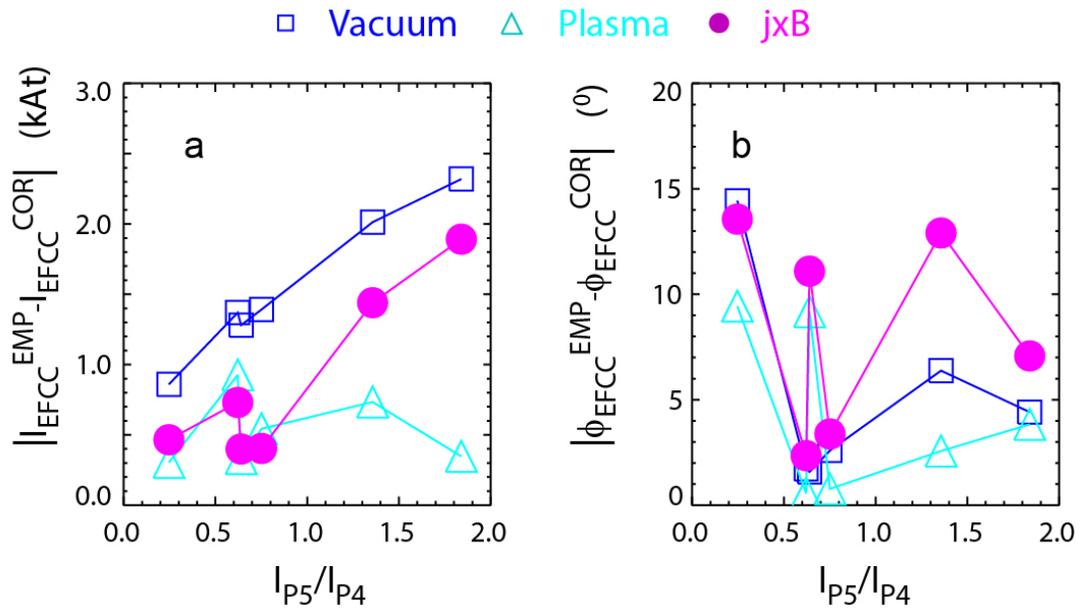

**Figure 19** Difference in a) amplitude and b) phase between the empirically derived correction current and criteria to minimise the n=1 m=2 radial field component at the q=2 surface using the vacuum approximation (square) or taking into account the plasma response (triangle) and minimising the jxB torque on the plasma (circle).